\documentstyle[twoside,fleqn,espcrc2,epsf,epsfig]{article}

\title{Infrared Abelian Dominance and Dual Higgs Mechanism 
in MA Gauge}
\author{H. Suganuma\address{Research Center for Nuclear 
Physics (RCNP), Osaka University, Ibaraki, Osaka 567-0047, 
Japan}\thanks{e-mail: suganuma@rcnp.osaka-u.ac.jp},
K. Amemiya$^{\rm a}$ and H. Ichie$^{\rm a}$ }

\begin{document}

\begin{abstract}
{\normalsize
We study infrared abelian dominance and the dual Higgs mechanism 
in the maximally abelian (MA) gauge 
using the lattice QCD Monte Carlo simulation.
In the MA gauge, the off-diagonal gluon phase tends to be random, 
and the off-diagonal gluon $A_\mu^\pm$ acquires the effective mass 
as $M_{\rm off} \simeq$ 1.2 GeV. 
From the monopole current in the MA gauge, 
we extract the dual gluon field $B_\mu$ and estimate 
the dual gluon mass as $m_B \simeq $ 0.5 GeV. 
The QCD-monopole structure is also investigated 
in terms of off-diagonal gluons. 
From the lattice QCD in the MA gauge,
the dual Ginzburg-Landau (DGL) theory can be constructed 
as a realistic infrared effective theory based on QCD.
}
\end{abstract}
\maketitle

\section{Strong Randomness of Off-diagonal Gluon Phase in MA gauge}

We find {\it strong randomness of the off-diagonal gluon phase} 
$\chi_\mu$ in the MA gauge using the SU(2) lattice QCD  
\cite{ichieNP,sugaPTP}.
In fact, we find very small correlation between neighboring phases, 
$\chi_\mu(s)$ and $\chi_\mu(s+\hat \nu)$, on the lattice 
in the MA and U(1)$_3$ Landau gauge.
This tendency seems natural because of the following reasons. 
Off-diagonal gluon phase $\chi_\mu$ is not constrained 
by the MA gauge-fixing condition at all, 
and the constraint from the QCD action is also small because 
of strong suppression of 
the off-diagonal gluon amplitude $|A_\mu^{\pm}|$ 
in the MA gauge. 
Within the random phase-variable approximation, 
{\it perfect abelian dominance for the string tension} can be 
clearly demonstrated \cite{ichieNP,sugaPTP}.

\section{Propagator and Effective Mass of \\
Off-diagonal Gluon in MA gauge} 

Using the SU(2) lattice QCD with $12^3 \times 24$, $16^4$ and $20^4$,  
we study the off-diagonal gluon propagator   
$G_{\mu\mu}^{\rm off} (r) \equiv \langle A_\mu^+ (x)A_\mu^- (y) \rangle$  
as the function of $r \equiv \sqrt{(x-y)^2}$ in the MA gauge 
with U(1)$_3$ Landau gauge fixing \cite{sugaPTP,ameNP}. 
From the slope of the logarithmic plot of $r^{3/2} G_{\mu\mu}^{\rm off}(r)$ 
in Fig.1(a), we get the off-diagonal gluon mass as 
$M_{\rm off}\simeq$ 1.2 GeV in the MA gauge \cite{sugaPTP,ameNP} 
as shown in Fig.1(b). 
Thus, in the MA gauge, the off-diagonal gluon cannot carry 
the long-range interaction such as the confinement force.
This is the essence of infrared abelian dominance.
In fact, QCD in the MA gauge behaves as a strong-coupling 
compact QED at the larger scale than $M^{-1}_{\rm off}\simeq$ 0.2 fm 
\cite{sugaPTP,ameNP}.

\begin{figure}[htb]
\vspace{-1.4cm}
\begin{center}
\epsfig{figure=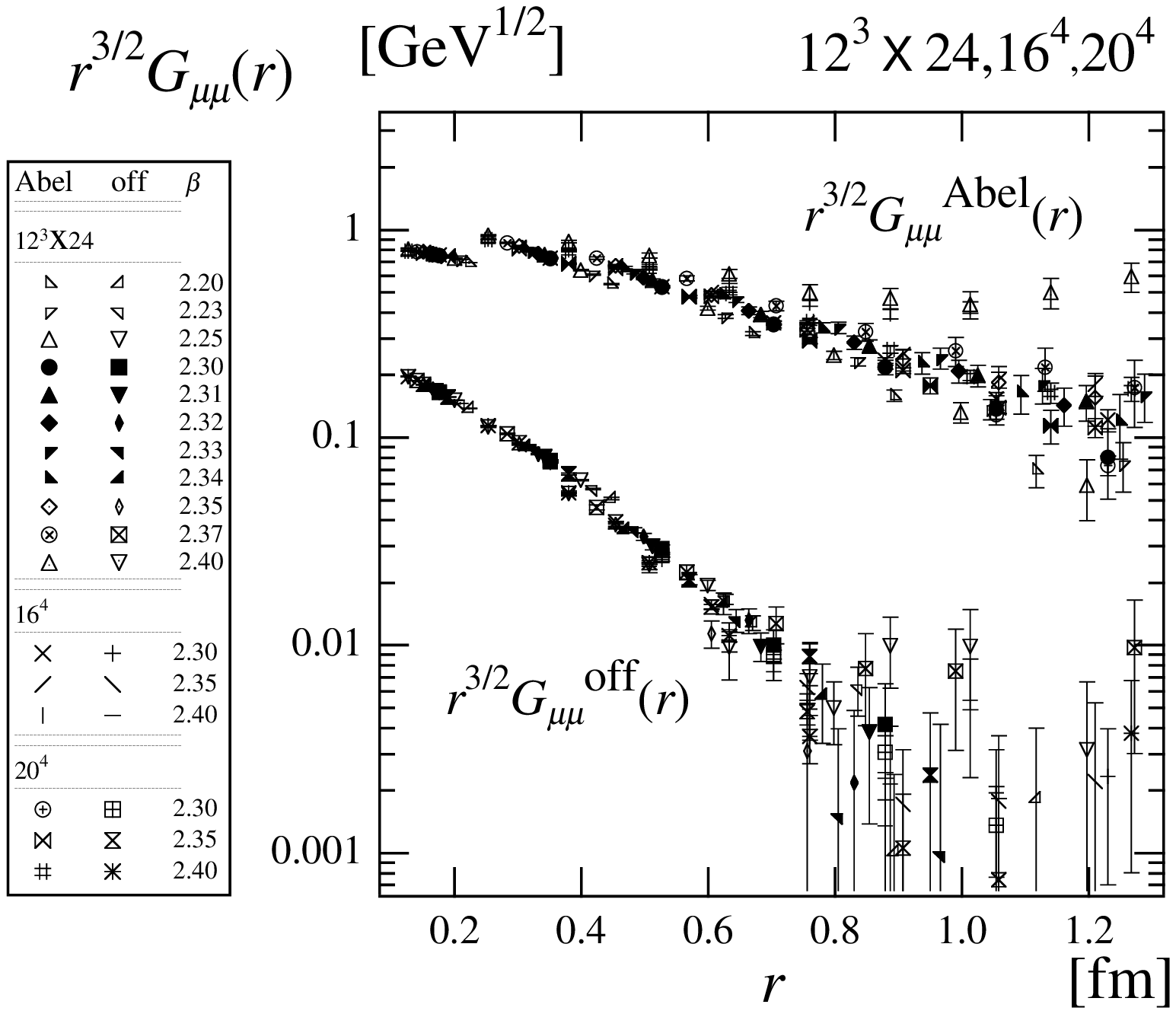,height=4.5cm}
\vspace{-1.1cm}
\epsfig{figure=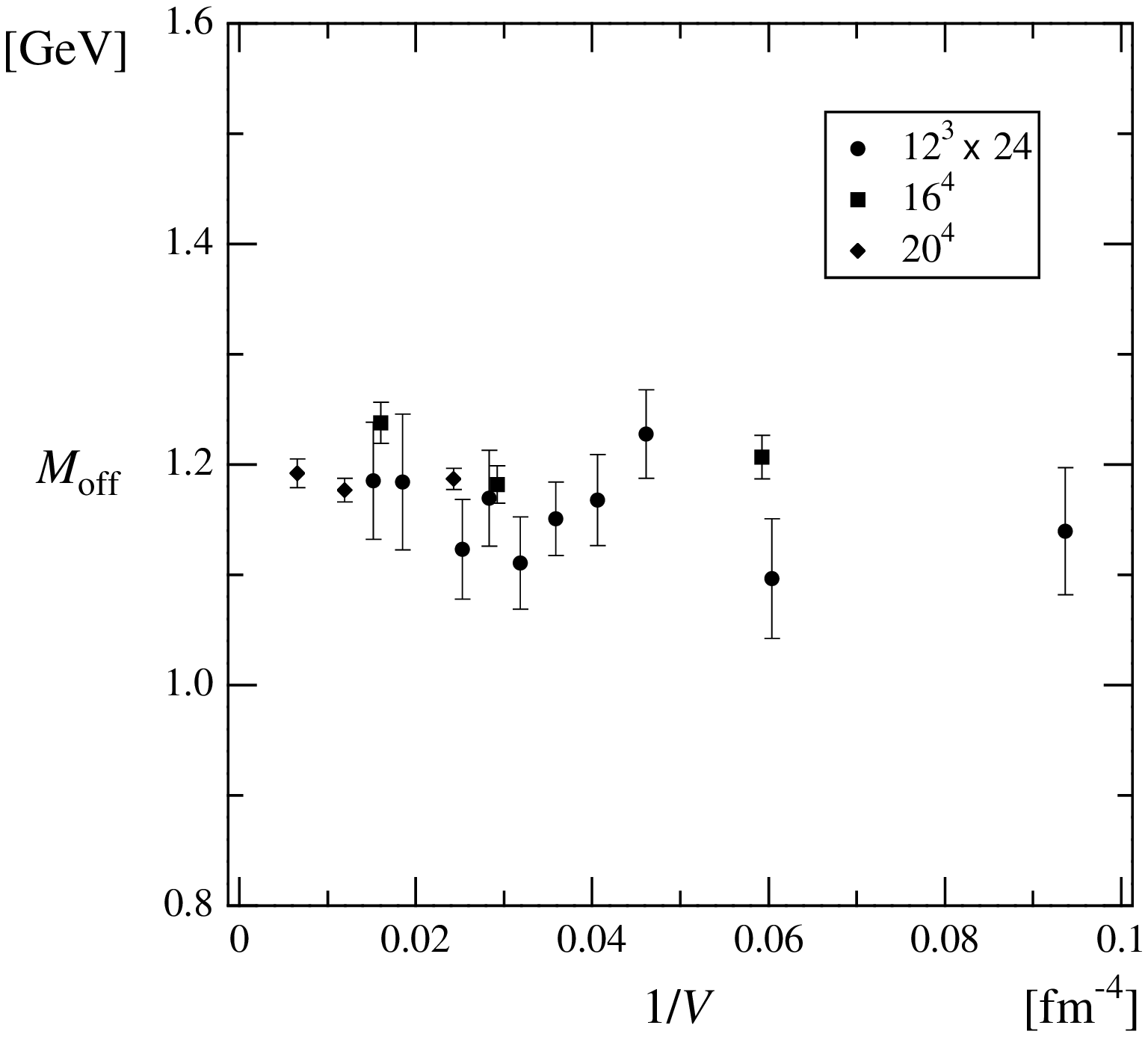,height=3.6cm}
\caption{
(a) The logarithmic plot of  
$r^{3/2} G_{\mu \mu}^{\rm off}(r)$, 
$r^{3/2} G_{\mu \mu}^{\rm Abel}(r)$. 
The slope corresponds to the effective mass.
(b) $M_{\rm off}$ v.s. inverse lattice volume. 
}\label{cap:offgluon}
\vspace{-1.5cm}
\end{center}
\end{figure}

\section{Dual Gluon Mass in MA gauge} 

Using the SU(2) lattice QCD with $20^4$ and $24^4$, 
we extract the dual gluon field $B_\mu$ from 
the monopole current $k_\mu$ appearing in the MA gauge, 
and investigate the dual gluon properties \cite{sugaPTP}. \\
(1) From the dual Wilson loop, 
we obtain the inter-monopole potential 
in the monopole part. \\
(2) The dual gluon propagator 
$\langle B_\mu (x) B_\mu (y) \rangle$ is measured 
in the dual Landau gauge $\partial_\mu B_\mu=0$. \\ 
As the evidence of the dual Higgs mechanism 
by color-magnetic monopole condensation, 
the dual gluon mass is estimated as $m_B \simeq$ 0.5 GeV 
from the lattice data in 0.4 fm $< r <$ 1.5 fm \cite{sugaPTP}.
Here, the dual gluon mass $m_B \simeq$ 0.5 GeV 
provides the key parameter of the DGL theory.

\begin{figure}[htb]
\vspace{-0.7cm}
\epsfig{figure=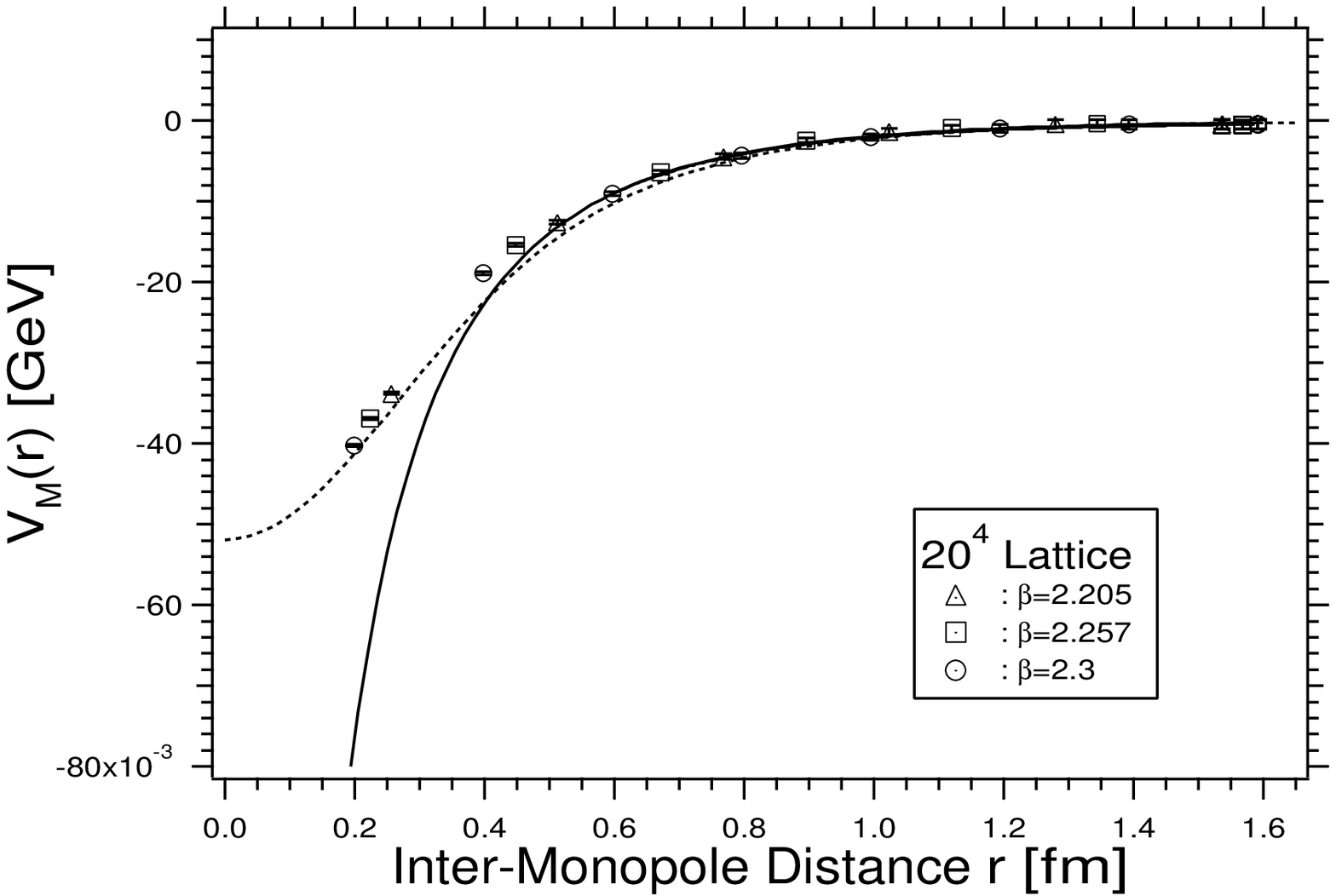,height=2.75cm}
\hspace{.1cm}
\epsfig{figure=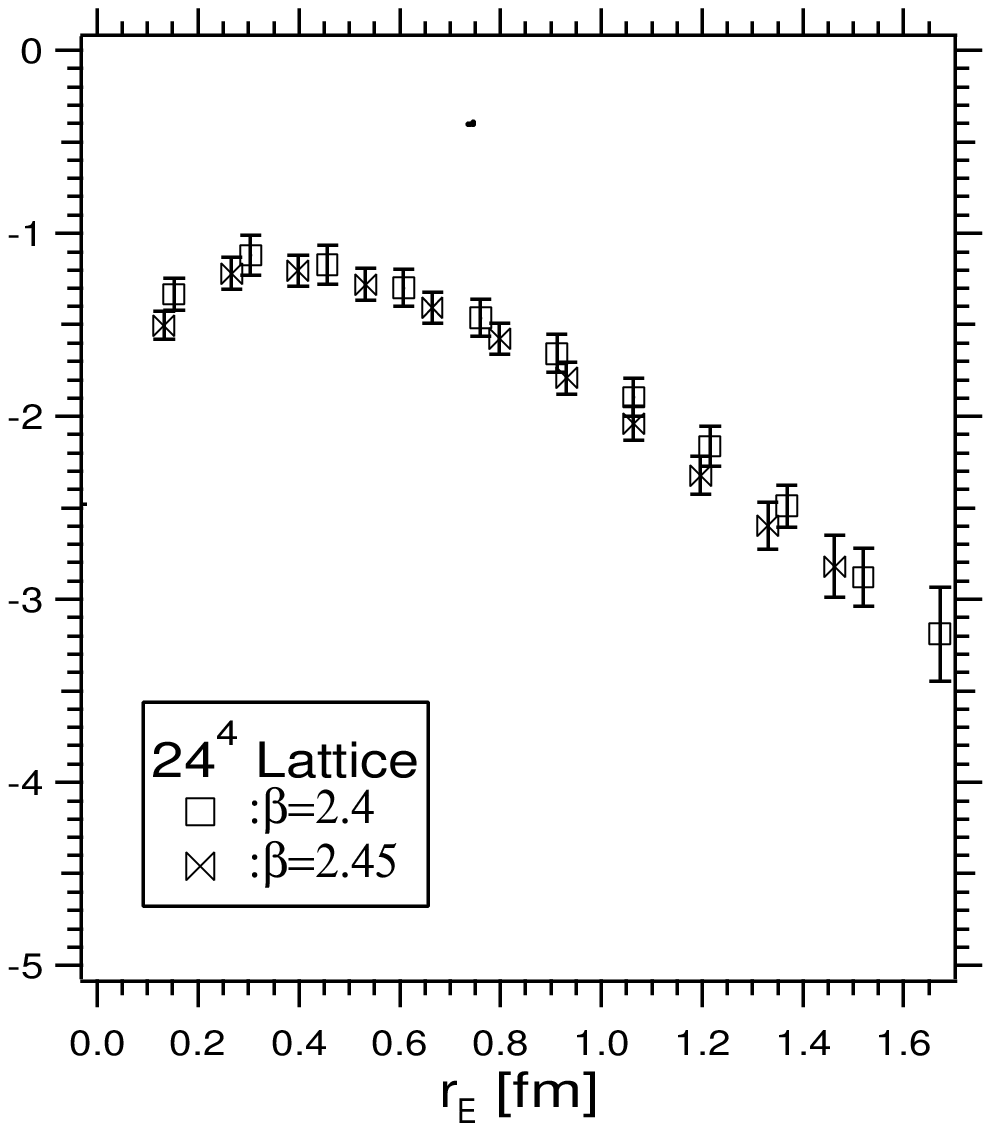,height=2.75cm}
\vspace{-1.0cm}
\caption{
(a) The inter-monopole potential. 
(b) $\ln (r_E^{3/2}\langle B_\mu(x)B_\mu(y) \rangle)$ 
v.s. 4 dim. distance $r_E$.
}\label{cap:dualgluon}
\vspace{-1.1cm}
\end{figure}

\section{Structure of QCD-Monopoles in terms of Off-diagonal Gluons}

We investigate the QCD-monopole structure in the MA gauge 
in the lattice QCD \cite{sugaPTP}.
\\
(1) Around the QCD-monopole, both the abelian action density 
$S_{\rm Abel}$ and the off-diagonal gluon contribution 
$S_{\rm off} \equiv S_{\rm QCD}-S_{\rm Abel}$ are largely fluctuated, 
and their cancellation keeps the total QCD action 
density $S_{\rm QCD}$ small \cite{sugaPTP}. 
\\
(2) The QCD-monopole has an intrinsic structure 
relating to a large amount of off-diagonal gluons around its center, 
similar to the 't Hooft-Polyakov monopole \cite{sugaPTP}. 
At a large scale where this structure becomes invisible, 
QCD-monopoles can be regarded as point-like Dirac monopoles.
\\
(3) From the concentration of off-diagonal gluons around QCD-monopoles 
in the MA gauge, we can naturally understand the {\it local 
correlation between monopoles and instantons.}
In fact, instantons tend to appear around the monopole world-line 
in the MA gauge, because instantons need full SU(2) gluon 
components for existence \cite{sugaPTP,fukuPR}. 

\section{Construction of the DGL theory from the lattice QCD in MA gauge}

In the MA gauge, the off-diagonal gluon contribution can be 
neglected and monopole condensation occurs at the infrared scale of QCD. 
Therefore, the QCD vacuum in the MA gauge can be regarded as 
the dual superconductor described by the DGL theory, 
and quark confinement can be understood with the dual Meissner effect.
The DGL theory can describe not only quark confinement \cite{sugaNP}
but also dynamical chiral-symmetry breaking (D$\chi$SB) 
by solving the Schwinger-Dyson equation for 
the quark field \cite{sugaNP,sasaki}. 
Monopole dominance for D$\chi$SB is also found from 
the analysis of the effective potential \cite{umi}. 
Using the DGL theory, we have studied 
the QCD phase transition (deconfinement \cite{sugaNP,ichiePR}
and chiral restoration \cite{sasaki}) at finite temperature, 
the QGP creation process in ultra-relativistic 
heavy-ion collisions \cite{sugaNP,ichiePR}, 
hadronization in the early universe \cite{mon},  
and glueball properties \cite{koma} systematically.

\begin{figure}[htb]
\center{
\epsfig{figure=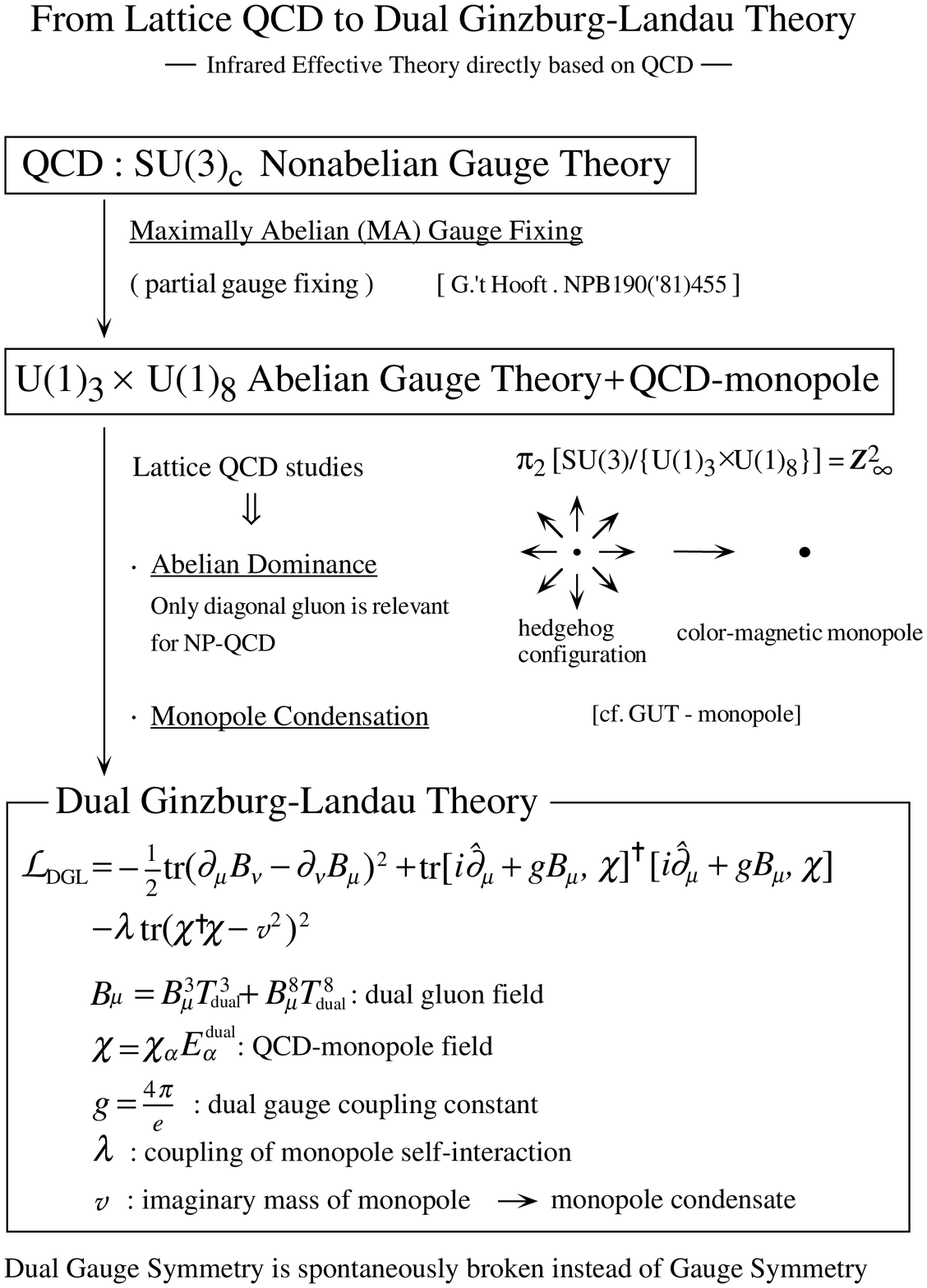,height=20.cm}
}
\vspace{-1cm}
\caption{
Construction of the dual Ginzburg-Landau (DGL) theory 
from the lattice QCD in the maximally abelian (MA) gauge.}
\end{figure}

\end{document}